\documentclass{desyproc}

\begin{document}
%------------------------------------
\title{Fermi-LAT and NuSTAR as 
	Stellar Axionscopes
}
\author{{\slshape Maurizio Giannotti}\\[1ex]
	Barry University, Miami Shores, US}

\contribID{familyname\_firstname}

\confID{13889}  
\desyproc{DESY-PROC-2017-XX}
\acronym{Patras 2017} 
\doi  

\maketitle

\begin{abstract}
We overview the potential of airborne X-ray and gamma-ray telescopes to probe the axion-like particles parameter space.  
\end{abstract}

\section{Introduction}

Besides contributing to stellar cooling, axion-like particles (ALPs) produced in stars could decay or, in the presence of an external magnetic field, oscillate into photons, providing a way for direct detection. 
This strategy is currently adopted by experiments such as CAST~\cite{Anastassopoulos:2017ftl} and NuSTAR~\cite{Harrison:2013md} to look for solar axions.

Other stars could contribute as well to a potentially observable photon flux detectable on Earth with current space born X-ray and gamma-ray instruments.  
Interestingly, in certain regions of the ALP parameter space such instruments, developed with the purpose of studying high energy photons, not ALPs, could exceed the probing potential of dedicated experiments such as ALPS II~\cite{Bahre:2013ywa} and IAXO~\cite{Armengaud:2014gea,Giannotti:2016drd}.

\section{Light ALPs form Supernovae}

Of particular interests for the study of ALPs are supernova (SN) events. 
In the extreme conditions of the SN core, ALPs can be efficiently produced and, given the small couplings we are interested in, stream freely out of it.
The total production rate of light ALPs ($ m_a< T$) per unit energy (integrated over the explosion time) can be approximated as\footnote{Here we are assuming that ALPs interact only with photons. In this case, they can be produced in the SN core through the Primakoff process~\cite{Payez:2014xsa}, in which the ALP converts into a photon in the proton electrostatic field. It is, however, also possible that ALPs interact with nuclei. In particular, standard QCD axions do. In this case, the nuclear Bremsstrahlung may be more efficient. In general, the two spectra produced by these processes are similar in shape, with the Nuclear Bremsstrahlung peaked at a slightly lower energy. } 
\begin{equation}
\frac{dN_a}{dE} \simeq C 
\,g_{12}^{2}
\left(E/E_0\right)^\beta e^{-(\beta + 1) E/E_0} \,, 	
\label{eq:time-int-spec}
\end{equation}
with $ g_{12}= g_{a\gamma}/(10^{-12}{\rm GeV}^{-1})$.
The other parameters depend on the progenitor mass.
For a progenitor of 10-18$ M_\odot $, one finds
  $ C \simeq (5-9)\times 10^{48}$ MeV$ ^{-1} $ 
  $ E_0 \simeq 100$ MeV, and $ \beta \simeq 2$~\cite{Payez:2014xsa,Meyer:2016wrm}.
Notice that the average ALP energy is approximately $ 3E_0(1+\beta)^{-1} \simeq 100 $ MeV  and the spectrum is maximal for $ E\simeq 2E_0(1+\beta)^{-1} \simeq 60 $ MeV. 
Integrating over the energy, we find a production of a few $ 10^{49} g_{12}^{2}$ ALPs over the time of the SN explosion.

If light enough, some of these ALPs oscillate into photons in the external magnetic field, producing a flux $ F_\gamma $ on earth
%.......................................................
\begin{equation}
\frac{dF_\gamma}{dE}= \frac{1}{4 \pi d^2} \frac{dN_a}{dE} \times P_{a\gamma} \,,\label{eq:diffphotonflux}
\end{equation}
%.......................................................
where $ P_{a\gamma} $ is the oscillation probability and $d$ the SN distance.

Since (for sufficiently light ALPs) the energy dependence in $ P_{a\gamma} $ drops for photons of energy above 10 MeV or so~\cite{Payez:2014xsa}, the photon spectrum resembles the ALP spectrum.
Thus, one expects a photon flux on earth with average energy $ \sim $ 100 MeV, peacked at about 60 MeV.  
The ideal instruments to study the (ALP-induced) SN photon flux are, therefore, detectors sensitive to gamma rays in the energy band between a few MeV and a few 100 MeV.

\begin{figure}[t]
	\centerline{
		\includegraphics[width=0.45\textwidth]{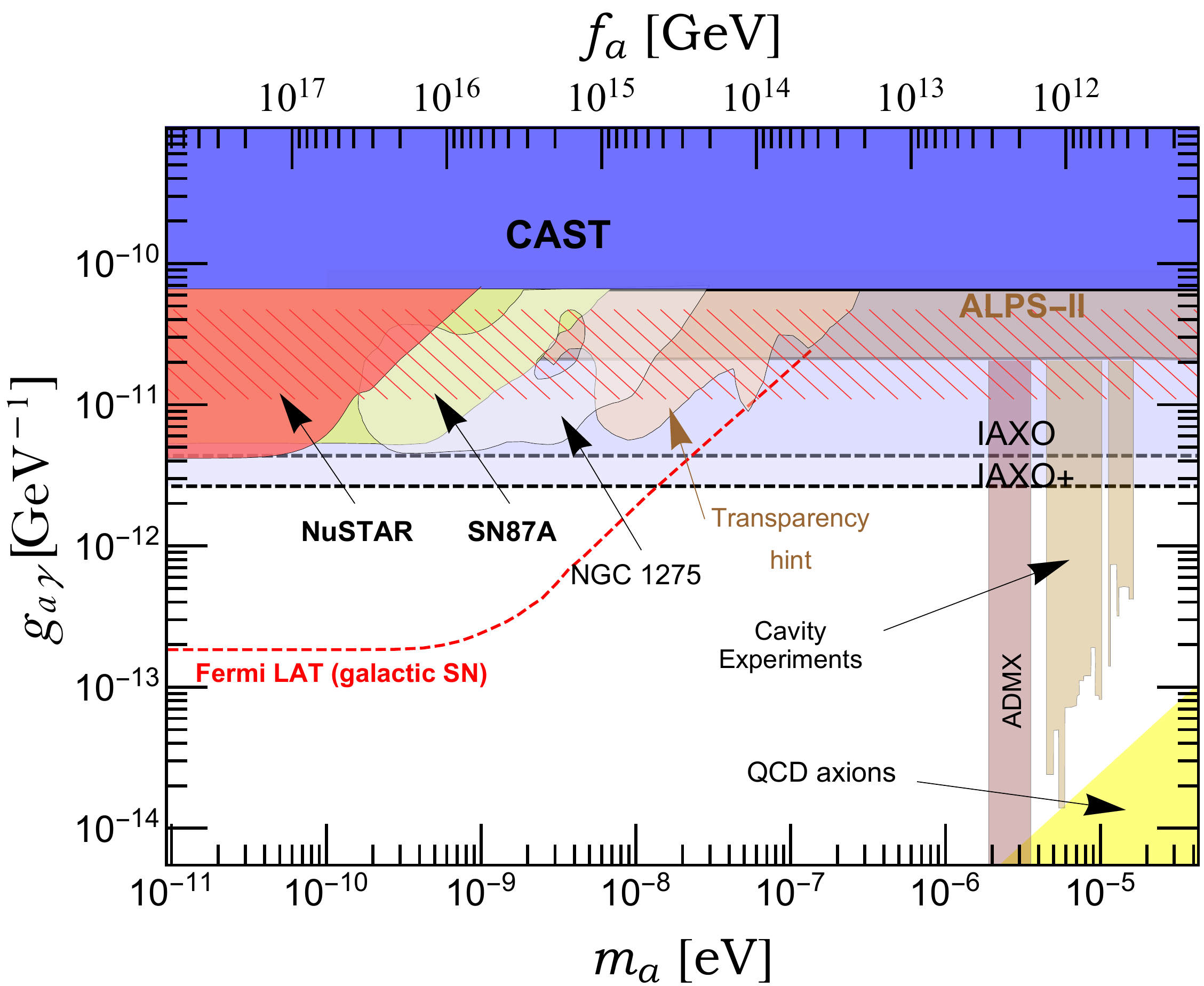}
		\hspace{0.5cm}
		\includegraphics[width=0.45\textwidth]{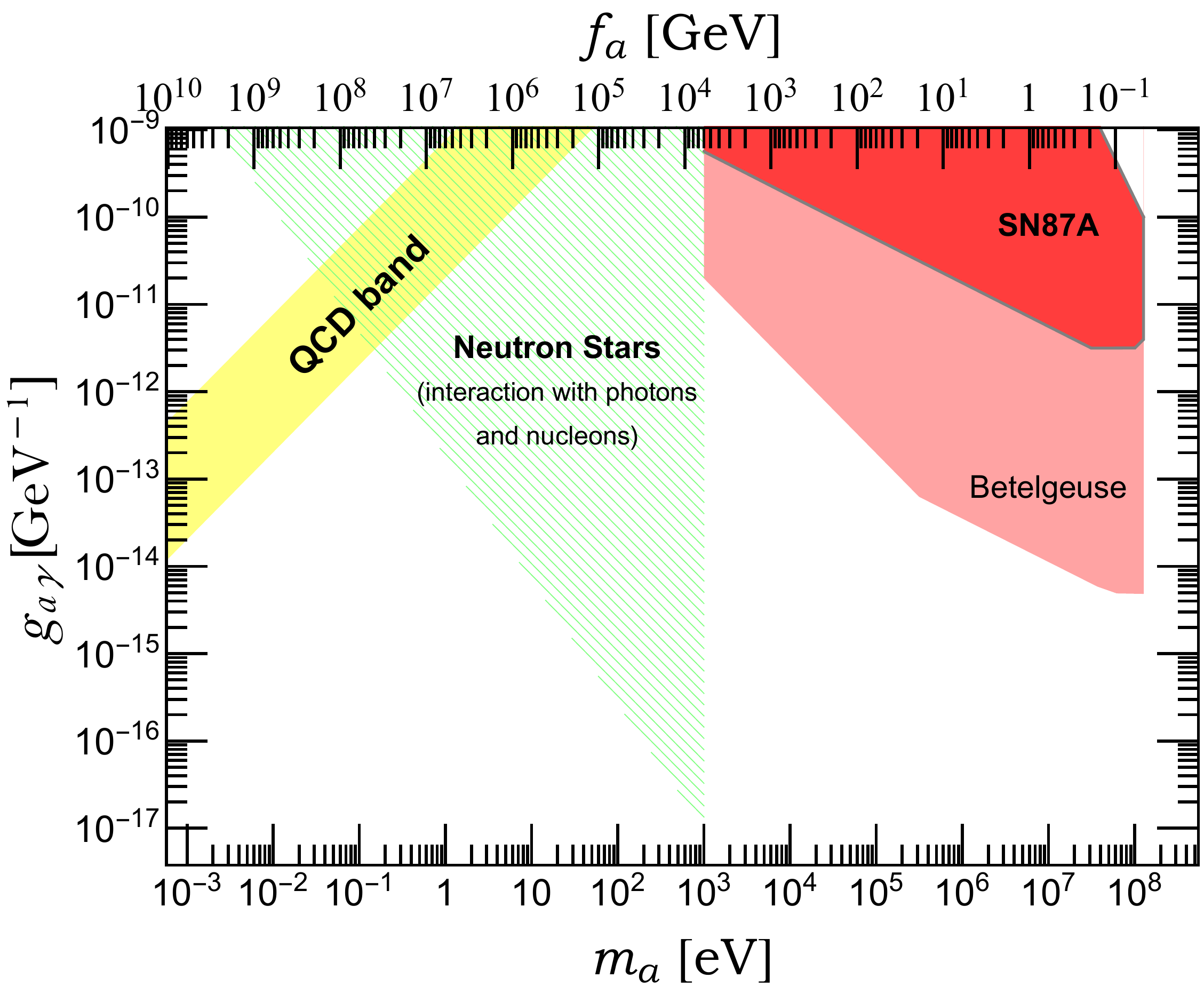}}
	\caption{\textit{Left panel:} Low mass ALPs parameter space.
		The hash-dashed region indicates the range of parameters expected to explain the stellar cooling anomalies~\cite{Ayala:2014pea,Giannotti:2015kwo}, which superimpose partially over the transparency hint region, invoked to explain the transparency of the universe to very high energy gamma rays~\cite{Meyer:2013pny}.
		This region is partially excluded by the search for spectral irregularities in the gamma ray spectrum of NGC~1275~\cite{TheFermi-LAT:2016zue}.
		The dashed red line is the expected Fermi potential for a next galactic SN~\cite{Meyer:2016wrm} discussed in the text.
		\textit{Right panel:} Bounds on heavy ALPs from NS and SN. Notice that the bound from the NS assumes interaction with both neutrons and photons. 
		The region assumes $ m_a < 2 m_e$, so that the decay channel into $ e^+\,e^- $ is forbidden. 
		The regions labeled SN1987A and Betelgeuse are extracted from ref.~\cite{Jaeckel:2017tud}.
		The mass region between 1 and 10 keV is not shown in the original literature and has been extrapolated in this plot. 
	}\label{Fig:gag_plane}
\end{figure}

Unfortunately, SN events close enough to be analyzed are fairly rare and 
presently, the only SN event we can consider to efficiently constrain the ALP-photon coupling is SN1987A.
In this case we have to rely on the old and poorly known Gamma-Ray Spectrometer (GRS) on the Solar Maximum Mission (SMM).
The effective area of the GRS, about 100 cm$ ^2 $ in the energy range 10-100 MeV,\footnote{Note, however, that there is little information about the response of this instrument in the literature. Moreover, the instrument was looking at the sun at the time of the explosion and so could have detected only off-axis SN photons. } 
is quite small, compared to the newer instruments.
Nevertheless, the absence of any photon excess observed by this instrument at the time of the neutrino burst from SN1987A 
%which provided a 3$ \sigma $ upper bound on the observed fluence of $ 0.6 \gamma \,$cm $ ^{-2} $ in the energy band 25--100 MeV~\cite{Chupp:1989kx}, 
allows to set  a stringent bound (comparable to the IAXO potential in that region) on the axion-photon coupling, $ g_{a\gamma}\lesssim~5\times~\!\!10^{-12}~$GeV$ ^{-1} $, for masses  $ m_{a} \lesssim 4.4\times 10^{-10} $~eV~\cite{Payez:2014xsa} (see green shaded area in the left panel of Fig.~\ref{Fig:gag_plane}). 

The Fermi Large Area Telescope (Fermi LAT) is currently the ideal candidate to probe axions in the event of a next galactic SN explosion, though it performs considerably better at higher energies $ E\gtrsim 1$ GeV. 
Its effective area at energies above 1 GeV is about 1~m$ ^{2} $ but the effective area averaged over the ALP spectrum is only 5500 cm$ ^2 $.

For a galactic SN, one finds $ d\lesssim 10 \,$kpc and $ B \simeq $ a few  $ \mu$G.
With these conditions, assuming massless ALPs with energies higher than a few 10 MeV and an axion-photon coupling $ g_{a\gamma}~\sim~10^{-13}-10^{-10} $~GeV$ ^{-1} $, one finds 
$  P_{a\gamma}\simeq (g_{a\gamma}B_Td)^2/4 $, where $ d $ is the SN distance and $ B_T $ the component of the magnetic field orthogonal to the photon beam~\cite{Payez:2014xsa}.
Interestingly, in these conditions the distance dependence in Eq.~\eqref{eq:diffphotonflux} drops. 
This approximation is valid when the SN distance is much smaller than the coherence length of the galactic magnetic field, $ l\sim 10\, $ kpc.
In the case of more distant sources, the flux is reduced because of the average over several magnetic domains.
As a rough estimate, we should expect $ F_\gamma \simeq g_{12}^4 {\rm cm}^{-2} $ for $ d\ll l $, where we have integrated Eq.~\eqref{eq:diffphotonflux} over all the frequencies.
Using this rough estimate and the average effective area, one should therefore expect about $ 5\times 10^3\,g_{12}^{2} $ events in the Fermi LAT, in case of a galactic SN explosion.
Assuming 5 to 7 events for a 2 $ \sigma $ signal, (the exact number depends on the background and is not necessarily the same for all SNe~\cite{Meyer:2016wrm}), one finds that Fermi LAT has the potential to probe the axion coupling down to $ g_{a\gamma} \sim$ a few $ 10^{-13} $ GeV$ ^{-1} $ for a galactic SN.
The accurate result~\cite{Meyer:2016wrm} is shown in Fig.~\ref{Fig:gag_plane} and is quite impressive. 
In the hypothesis of a future galactic SN during the time of the Fermi mission, we would be able to explore a region of the ALP parameter space with a very rich phenomenology, including the region of the transparency hints~\cite{Meyer:2013pny} and part of the region invoked for the stellar cooling anomalies~\cite{Giannotti:2015kwo,Giannotti:2017hny}.

Next generation instruments, such as e-Astrogram~\cite{DeAngelis:2016slk} and ComPair~\cite{Moiseev:2015lva}, are       especially efficient at lower energies and, more importantly, have a much smaller point spread function with respect to Fermi LAT at the energies expected from SN events. 
However, their effective areas averaged over the expected photon spectrum is about a factor of 4-5 smaller than the Fermi LAT one.
Although we have not performed a detailed analysis of the response function of these new instruments, it is unlikely that they would improve substantially (if at all) on the Fermi potential.
A full analysis of the potential of the next generation of gamma ray observatories for galactic SN is in preparation.

\section{Massive ALPs from Supernovae and Neutron Stars}

The interest in massive ALPs has increased in last few years thanks to improvements in the experimental potential to probe them~\cite{Dobrich:2017gcm,Dolan:2017osp} (see also B. Dobrich contribution to these proceedings). 
Non-minimal QCD axion models, such as those discussed in~\cite{Berezhiani:2000gh,Gianfagna:2004je,Dienes:1999gw}, predict massive ALPs, which interact with photons and standard model fermions. 
If massive enough, ALPs could \textit{decay} rather than oscillate into photons, and produce a sufficient photon flux on earth, regardless of the magnetic field.  
%This mechanism does not require an external magnetic field and is very efficient for masses around 1 MeV.

The phenomenology of photons from massive ALPs from astrophysical sources is quite interesting. 
In particular, the arrival time of these photons could, in general, be much longer than the typical explosion time~\cite{Giannotti:2010ty,Jaeckel:2017tud}. 
In studying the experimental potential it is therefore necessary to account for longer detection times.

A thorough analysis of the constraints on ALPs from SN1987A was presented in~\cite{Jaeckel:2017tud}, where the authors also estimated the Fermi LAT potential in case the close red supergiant Betelgeuse were to go SN.
The result is shown in the right panel of Fig.~\ref{Fig:gag_plane}.
The analysis assumes ALPs interacting with photons only.
The figure shows also the Fermi exclusion plot derived from the analysis of 5 years of gamma-ray data for a sample of 4 nearby neutron stars (NS)~\cite{Berenji:2016jji}.
Notice, however, that the region shown is model dependent and assumes axions interacting not only with photons but also with neutrons, with $ g_{a\gamma}\simeq 10^{-2} g_{an}$~GeV$ ^{-1} $.
In fact, in the proton-poor NS environment the Primakoff production process is very modest and the only efficient axion production mechanism is the neutron bremsstrahlung.

\section{Betelgeuse}

One of the most interesting stars to study ALPs, besides our sun, is Betelgeuse~\cite{Carlson:1995wa}, a supergiant in the  constellation of Orion, about $ 200 \, $pc from the sun. 
Though not the closest star to our sun, Betelgeuse has a much higher core temperature than nearer stars and would therefore be a better source of ALPs. 

Light ALPs can be efficiently produced in the star through the Primakoff process, which requires interaction with photons only, and then converted into photons in the galactic magnetic field.
Using $ B_T=2.9~\mu $G, $ d=197\, $pc, and $ m_a=0 $, we find that the expected photon spectrum on earth is well approximated by
\begin{equation}
\frac{dN_\gamma}{dE\,dt} \simeq C 
%\left(\frac{g_{a\gamma}}{10^{-11}{\rm GeV}^{-1}}\right)^2
\,g_{10}^{2}
\left(E/E_0\right)^\beta e^{-(\beta + 1) E/E_0} \,, 
\label{eq:spec}
\end{equation}
with $ C\simeq (0.4-1) $ keV$ ^{-2} $ cm$ ^{-2} $ s$ ^{-1} $, $ \beta\simeq 2 $, $ E_0\simeq 60-100 $ keV.
The coefficients depend on the stellar model, which is the greatest source of uncertainty. 

This mechanism would produce a photon flux peaked in the hard X-ray region, at about 50-60 keV. 
Most X-ray detectors are not very efficient in this region.
Currently, the most efficient is NuSTAR~\cite{Harrison:2013md}, though its effective area is steeply reduced at energies above a few 10 keV.
Integrating over the effective area in~\cite{Harrison:2013md}, one can expect 
$ \sim 300 \,g_{10}^2$ photons from ALP conversion per second in NuSTAR, a value considerably larger than 
$ \sim 3 \,g_{10}^2 \, {\rm s}^{-1} $ photons in Chandra or 
$ \sim 2 \,g_{10}^2 \, {\rm s}^{-1} $ photons in XMM-Newton.
Assuming a background $ \simeq 10^{-3} \,\gamma \, {\rm s}^{-1} $~\cite{Harrison:2013md}, one should expect NuSTAR to be able to probe values of the axion-photon coupling down to a few $ 10^{-12} $ GeV$ ^{-1} $, the same level of sensitivity expected by IAXO. 
An estimate of the NuSTAR sensitivity is shown in the left panel in  Fig.~\ref{Fig:gag_plane}.
A more comprehensive and detailed analysis is in preparation. 
%In Fig.~\ref{Fig:gag_plane}, we show a preliminary estimate of what the expected potential could be. 

Regardless of the precise level of sensitivity, it is in general obvious that NuSTAR observations of Betelgeuse would allow to probe couplings at least as low as those reached by the SN1987A analysis, for masses below $ m_a\sim$ a few $ 10^{-11} $ eV, without being subject to the same level of uncertainties. 

\section*{Acknowledgments}
I would like to thank B. D\"obrich, B. Grefenstette, M. Meyer, A. Mirizzi, K. Perez, J. Ruz Armendariz, O. Straniero and J. K. Vogel for useful discussions and suggestions. 
A special thank to the organizers of this very interesting workshop. 
% ****************************************************************************
% BIBLIOGRAPHY AREA
% ****************************************************************************

\begin{footnotesize}

\end{footnotesize}

% ****************************************************************************
% END OF BIBLIOGRAPHY AREA
% ****************************************************************************

\end{document}